\documentclass[12pt,fleqn,a4paper]{article}
\usepackage{epsfig,amssymb,amsthm,amsmath,mathrsfs}

\newcommand{\rem}[1]{} 

\newcommand{\R}{{\mathbb{R}}}

\newcommand{\Z}{{\mathbb{Z}}}

\newcommand{\res}{\mathop{\mathrm{Res}}}

\newcommand{\eps}{\varepsilon}
\newcommand{\imag}{\mathrm{i}}

\newcommand{\dee}{{\rm d}}
\newcommand{\sgn}{{\rm sgn}}

\newtheorem{theorem}{Theorem}[section]
\newtheorem{lemma}[theorem]{Lemma}
\newtheorem{proposition}[theorem]{Proposition}
\newtheorem{corollary}[theorem]{Corollary}
\newtheorem{definition}[theorem]{Definition}

\begin{document}

\title{Geodesics on the Ellipsoid and Monodromy}

\author
{
{\protect\normalsize Chris M. Davison, Holger R.\ Dullin, Alexey V.\ Bolsinov} \\
{\protect\footnotesize\protect\it
Department of Mathematical Sciences, Loughborough University} \\[-2mm]
{\protect\footnotesize\protect\it
Leicestershire, LE11 3TU, United Kingdom\footnote{ email:
c.m.davison@lboro.ac.uk, h.r.dullin@lboro.ac.uk, a.bolsinov@lboro.ac.uk
% phone: +44(0)150922-3187, fax -3969
}.}
}

\date{\protect\normalsize 5 September 2006}

\maketitle

\begin{abstract}
The equations for geodesic flow on the ellipsoid are well known, and were first solved by Jacobi in 1838 by separating the variables of the Hamilton-Jacobi equation.  In 1979 Moser investigated the case of the general ellipsoid with distinct semi-axes and described a set of integrals which weren't know classically.  After reviewing the properties of geodesic flow on the three dimensional ellipsoid with distinct semi-axes, we investigate the three dimensional ellipsoid with the two middle semi-axes being equal, corresponding to a Hamiltonian invariant under rotations.  The system is Liouville-integrable and thus the invariant manifolds corresponding to regular points of the energy momentum map are 3-dimensional tori.  An analysis of the critical points of the energy momentum maps gives the bifurcation diagram.  We find the fibres of the critical values of the energy momentum map, and carry out an analysis of the action variables.  We show that the obstruction to the existence of single valued globally smooth action variables is monodromy.
\end{abstract}

\section{Introduction}

The geodesic flow on the ellipsoid is the classical example of a non-trivial separable and thus Liouville 
integrable Hamiltonian system. It is the prime example in Jacobi's famous 
``Vorlesungen \"uber Dynamik'' \cite{jacobi84} and may be considered as his motivation to develop 
Hamilton-Jacobi theory and the solution of the Abel-Jacobi inversion problem.
Its modern treatment was pioneered by the Z\"urich school, 
namely by Moser \cite{moser80} and Kn\"orrer \cite{knorrer80,knorrer82}, generalising to the $n$-ellipsoid and providing
smooth integrals and the general solution in terms of $\theta$-functions
for the generic case of an $n$-ellipsoid with pair-wise distinct semi-axes.
Separation leads to a curve of genus $n$ and the $n$ actions are given by 
integrating a differential of second kind over a basis of real cycles.
The generic motion on an $n$-torus corresponds to a non-degenerate curve. Special 
motions correspond to degenerate curves. 
For two degrees of freedom the topology of the Liouville foliation was analysed
in \cite{bolsinov95}.
An excellent general approach to the topology of St\"ackel systems, including geodesic flow
on the ellipsoid with distinct semi-axes, was carried out by Zung \cite{zung96}.  This is the first
paper where the singularities of the Liouville foliation were studied for geodesic flow on the
ellipsoid.  We extend his results to the degenerate case.
%It shows that the image the the energy-momentum map is a disjoint 
%union of simply connected chambers of regular values. 

Surprisingly most of these results are not stable when the ellipsoid becomes degenerate, 
i.e.\ when some semi-axes coincide. The smooth integrals develop poles in 
this limit, the hyperelliptic curve changes, complete Abelian actions integrals change from 2nd kind
to 3rd kind, 
and the topology of the critical values in the image of the energy-momentum map
changes. Here we study these changes for the 3-ellipsoid.
The most interesting result appears when the middle-axes coincide; the set of regular values of the energy-momentum map becomes non-simply connected.  Duistermaat  \cite{duistermaat80} realised that in this case global action variables might not exist.  The torus-bundle over the regular values is non-trivial and has monodromy.  This shows that by making the system simpler (i.e. more symmetric) it can become more complicated (i.e. have a non-trivial torus bundle).

\section{The geodesic flow on generic 3-ellipsoids}

A 3-ellipsoid  embedded in $\R^4$ with 
coordinates $x = (x_0, x_1, x_2, x_3)$  has the equation 
$\langle  A^{-1} x , x\rangle = 1$ with the standard Euclidean scalar 
product $\langle, \rangle$ and a positive definite matrix $A$. 
This quadratic form can always be diagonalised by an orthogonal transformation
and the eigenvalues of $A$ are denoted by $0 < \alpha_0 \le \alpha_1\le \alpha_2 \le \alpha_3 $.
Thus we may assume that $A$ is diagonal and the standard form of the ellipsoid with semi-axis 
$\sqrt{\alpha_i}$ embedded in $\R^4$ is
\[
C_1 = 
    \frac{x_0^2}{\alpha_0} + 
    \frac{x_1^2}{\alpha_1} + 
    \frac{x_2^2}{\alpha_2} + 
    \frac{x_3^2}{\alpha_3}  
    - 1 = 0 \,.
\]
For the generic non-degenerate ellipsoid the semi-axes are distinct.
The Lagrangian of a free particle with mass 1 is 
$L = \frac12( \dot x_0^2 +  \dot x_1^2 +  \dot x_2^2 + \dot x_3^2 )$.
The equations of motion with Lagrange multiplier $\Lambda$ enforcing the 
constraint $C_1 = 0$ are 
\begin{equation} \label{Newton}
   \ddot x = -\Lambda A^{-1} x, \qquad 
   \Lambda = \frac{\langle A^{-1} \dot x, \dot x \rangle}{\langle A^{-1} x, A^{-1} x\rangle} \,.
\end{equation}
A Hamiltonian description can be obtained by introducing momenta $y_i = \dot x_i$
and enforcing the constraint by replacing the standard symplectic structure $\dee x \wedge \dee y$ 
by a Dirac bracket. 
The Dirac bracket has as Casimirs the constraint for being on the ellipsoid $C_1 = 0$ and the constraint for its tangent space
\[
C_2 = 
\frac{x_0 y_0}{\alpha_0} +
\frac{x_1 y_1}{\alpha_1} +
\frac{x_2 y_2}{\alpha_2} +
\frac{x_3 y_3}{\alpha_3}  = 0 \,.
\]
With the notation
\[
    D =
    \frac{x_0^2}{\alpha_0^2} + 
    \frac{x_1^2}{\alpha_1^2} + 
    \frac{x_2^2}{\alpha_2^2} + 
    \frac{x_3^2}{\alpha_3^2}  
    = \frac12 \sum \frac{\partial C_1}{\partial x_i} \frac{ \partial C_2}{\partial y_i}
\]
% and the angular momenta $L_{ij} = x_i y_j - x_j y_i$ 
the Dirac bracket with Casimirs $C_1$ and $C_2$ is given by 
\begin{equation}
\label{eqn:dirac}
\left\{x_i,x_j\right\} = 0, \qquad 
\left\{x_i,y_j\right\} = \delta_{ij} - \frac {x_ix_j} {D \alpha_i \alpha_j }, \qquad 
\left\{y_i,y_j\right\} = - \frac { x_i y_j - x_j y_i} {D \alpha_i \alpha_j} \,.
\end{equation}
The Hamiltonian is $H = \frac12( y_0^2 + y_1^2 + y_2^2 + y_3^2)$
and the equations of motion are
\begin{equation}
    \dot x_i = \{x_i,H\}, \qquad \dot y_i = \{y_i,H\}, \quad i = 0,1,2,3.
\end{equation}
These equations are equivalent to \eqref{Newton}.
The Hamiltonian vector field generated by $H$ is denoted by $X_H$.

The system is Liouville integrable with smooth global integrals 
(in the generic case of distinct semi-axes) first found by Uhlenbeck and Moser \cite{moser80}
\begin{equation}
F_i = y_i^2 + \sum_{j=1, j \ne i}^{n} \frac {\left(x_iy_j - x_jy_i\right)^2} {\alpha_i - \alpha_j}, 
\quad i =  0, \dots, 3 \,.
\end{equation}
On the symplectic leaf of the Dirac bracket given by $C_1= C_2 = 0$
they are related by $\sum F_i/\alpha_i = 0$ and they have pair-wise vanishing brackets
\cite{moser80}.
The integrals $F_i$ are related to the Hamiltonian by $H = \frac12( F_0 + F_1 + F_2 + F_3)$.

Ellipsoidal coordinates are local coordinates on the ellipsoid that separate 
the Hamiltonian. They are defined as the roots $\lambda$ of $K(x,x;\lambda)= 1$ where
\[
   K(x,y; \lambda) =\sum \frac{ x_i y_i}{\alpha_i - \lambda} \,. 
\]
The equations $K(x,x;\lambda_i)=1$ are linear in $x_i^2$ and can be easily solved to give
\begin{equation}  \label{x2is}
x_i^2 = \frac{B(\alpha_i)}{A'(\alpha_i)}, \quad
B(z) = \prod_{j=0}^3 (\lambda_j - z), \quad
A(z) = \prod_{j=0}^3 (\alpha_j - z)  \,.
\end{equation}
Because of  the poles in $K$ for fixed $x$ the 4 roots satisfy
\[
  \lambda_0 \le \alpha_0 \le \lambda_1 \le \alpha_1 \le \lambda_2 \le \alpha_2 \le \lambda_3 \le \alpha_3 \,.
\]
Fixing $\lambda_0 = 0$ gives a coordinate system $(\lambda_1, \lambda_2, \lambda_3)$ 
on the ellipsoid since $K(x,x; 0)-1 = C_1$.
The coordinate transformation to the new variables $\lambda_i$ 
and their conjugate momenta $p_i$ gives
\[
   H = 4 \sum_i \frac12 p_i^2 \frac{ \prod_j (\alpha_j - \lambda_i)}{ \prod'_j (\lambda_j - \lambda_i)}.
\]
The primed product excludes the vanishing term with $j=i$.
The geodesic flow on the ellipsoid $\lambda_0 = 0$ 
is described by the invariant subset given by $p_0 = \lambda_0 = 0$.
The variables can be separated by using the van der Monde matrix $(\lambda_i^{j-1})_{ij}$
as a St\"ackel matrix \cite{eisenhart34}. With separation constants $s_i$ where $s_3 = 2h$ and 
$s_0 = 0$ the separated equations are
\begin{equation} \label{PQU}
p_i^2 =- \frac{Q(\lambda_i)}{4 A(\lambda_i)}, \quad
Q(z) = 2 h z^3 + s_2 z^2 + s_1 z + s_0.
\end{equation}
The system separates on the hyperelliptic curve $w^2 = -Q(z) A(z)$ of degree 7, 
hence genus 3.
The relation between $F_i$ and the separation constants $s_i$ is 
determined by the residues of the identity
\begin{equation} \label{eqn:FandS}
  \sum_{i=0}^3 \frac{F_i}{z - \alpha_i } = \frac{ Q(z) } {A(z) } \,.
\end{equation}
In particular 
% $h = s_3 = \frac12 \sum f_i$, 
$s_2 = -\sum_{cycl} f_0(a_1 + a_2 + a_3)$,
$s_1 = \sum_{cycl} f_0(a_1 a_2 + a_1 a_3 + a_2 a_3)$, where $f_i$ denotes a value of $F_i$.

The ellipsoidal coordinates $\lambda_j$ only determine the squares of the $x_j$ 
and thus have singularities when $x_j = 0$. 
Smoother coordinates $\phi_i$ on a covering torus designed so that their cotangent lift 
cancels the singularities given by $A(z)$ in \eqref{PQU}
are defined by 
\begin{equation} \label{ellidef}
     \frac{\dee \phi_i}{\dee \lambda_i} = \frac{1}{2\sqrt{(-1)^i A(\lambda_i) } }, \quad \alpha_{i-1} \le \lambda_i \le \alpha_i, \quad i = 1,2,3 \,.
\end{equation}
This defines elliptic functions $\lambda_i(\phi_i)$ with modulus 
$k^2 = (\alpha_3-\alpha_2)(\alpha_1 - \alpha_0)/( (\alpha_3-\alpha_1)(\alpha_2 - \alpha_0))$ 
given by the cross ratios of the semi-axes squared for $i = 1, 3$ 
and with complementary modulus $\sqrt{1-k^2}$ for $i = 2$. 
The momenta conjugate to $\phi_i$ are denoted by $\hat p_i$.
In this coordinate system the squares of the new momenta are smooth functions 
$\hat p_i^2 = (-1)^{i+1}Q(\lambda_i(\phi_i))$. The Hamiltonian $H$ and the constants of motions from
separation $S_1$, $S_2$ in these coordinates are
\begin{eqnarray*}
H & = & 
   \frac{\hat p_1^2}{2\lambda_1 (\lambda_2 - \lambda_1) (\lambda_3 - \lambda_1)}
+ \frac{\hat p_2^2}{2\lambda_2 (\lambda_2 - \lambda_1) (\lambda_3 - \lambda_2)}
+ \frac{\hat p_3^2}{2\lambda_3 (\lambda_3 - \lambda_1) (\lambda_3 - \lambda_2)}, \\
S_1 & = & 
   \frac{\lambda_2 \lambda_3 \hat p_1^2}{\lambda_1 (\lambda_2 - \lambda_1) (\lambda_3 - \lambda_1)}
+ \frac{\lambda_1 \lambda_2 \hat p_2^2}{\lambda_2 (\lambda_2 - \lambda_1) (\lambda_3 - \lambda_2)}
+ \frac{\lambda_1 \lambda_2 \hat p_3^2}{\lambda_3 (\lambda_3 - \lambda_1) (\lambda_3 - \lambda_2)}, \\
S_2 & = & -
   \frac{(\lambda_2 + \lambda_3)\hat p_1^2}{\lambda_1 (\lambda_2 - \lambda_1) (\lambda_3 - \lambda_1)}
- \frac{(\lambda_1 + \lambda_3)\hat p_2^2}{\lambda_2 (\lambda_2 - \lambda_1) (\lambda_3 - \lambda_2)}
- \frac{(\lambda_1 + \lambda_2)\hat p_3^2}{\lambda_3 (\lambda_3 - \lambda_1) (\lambda_3 - \lambda_2)}.
\end{eqnarray*}
In these formulas each $\lambda_i$ represents the elliptic function $\lambda_i(\phi_i)$.

The $(\phi, \hat p)$ coordinate system still is not a global coordinate system on the 
cotangent bundle of the 3-ellipsoid (such global coordinates do not exist). 
It has singularities at the ``umbilical curve"\footnote{The term ``umbilical curve" is used in analogy to the umbillic points on the two dimensional ellipsoid - we don't know whether it has a differential-geometric characterisation.} determined by 
$\lambda_1 = \lambda_2 = \alpha_1$ contained in the $x_1 = 0$ plane
and
$\lambda_2 = \lambda_3 = \alpha_2$ contained in the $x_2 = 0$ plane.
Explicit formulas are obtained by inserting these conditions into \eqref{x2is},
where $\lambda_3$ or $\lambda_1$ become the curve parameter, respectively.
Hence the umbilic curves are coordinate lines on the respective sub-ellipsoid. 
The umbilic curves are thus 2 topological circles in the $x_2$-$x_3$-plane with 
$x_0 > 0$ or $x_0 < 0$ and 2 topological circles in the
$x_0$-$x_1$-plane with $x_3 > 0$ or $x_3 < 0$.
At these points in configuration space $H$ and $S_i$ are singular for
arbitrary momenta and a different coordinate system needs to be used.

\begin{figure}
\centerline{\includegraphics[width=8cm]{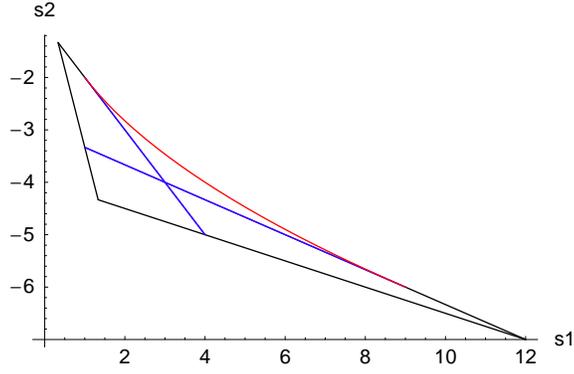}}
\caption{Bifurcation Diagram of the generic Ellipsoid with $\alpha_i = 1/3, 1, 3, 4$} 
\label{BifDiagGen}
\end{figure}

There are four 2-dimensional totally geodesic subflows obtained from 
setting $x_i = y_i = 0$. Similarly there are six 1-dimensional totally geodesic
subflows obtained from setting $x_i = x_j = y_i = y_j = 0$ for each pair of 
indices with $i < j$.  A 1-dimensional subflow is of course simply two periodic orbits. 
These six times two periodic orbits and the four subflows on 2-ellipsoids give the backbone of
the bifurcation diagram, i.e.\ the critical values in the image of the energy-momentum map.
These critical values are shown in figure~\ref{BifDiagGen}.

Each separated equation $\hat p_i^2 - (-1)^{i+1}Q(\lambda_i(\phi_i)) = 0$
defines a curve in the phase portrait in the $(\phi_i, \hat p_i)$ plane. 
The critical points occur at $\hat p_i = 0$ and
$\phi_i$ determined  by $Q'(\lambda_i) \lambda_i' = 0$.
Solutions are of two types: 
Either double roots of $Q$ or critical points of the elliptic function $\lambda_i$.
According to  \eqref{ellidef} critical points of $\lambda_i(\phi_i)$ occur exactly for 
$\lambda_i = \alpha_{i-1}$ or
$\lambda_i = \alpha_{i}$. 
By \eqref{x2is} this implies that $x_{i-1} = 0$ or $x_i = 0$, respectively,
and similarly for $y_i$ from the cotangent lift of \eqref{ellidef}.
These critical points therefore correspond to the geodesic subflows mentioned above.
The image of the critical points $x_j = y_j = 0$ is given by a segment of the 
line in the $s_1$-$s_2$ plane given by $Q(\alpha_j) = 0$, see figure~\ref{BifDiagGen}.
The other type of critical points occur for those values of $s_1, s_2$ for which 
there is a double root in $Q(z) =  2hz ( z - d)^2$ with $\alpha_1 \le d \le \alpha_2$ 
so that $\lambda_2 = d$ is fixed for this motion. For $d$ not at its boundary values
these critical points are {\em not} contained in any geodesic subflow.

Finally we have to establish whether any point with $x$-coordinates in the umbilic curves is critical.
The umbilic curves are contained in $x_1=0$ and $x_2=0$, respectively. 
The gradient of $F_i$ vanishes on the subflow $x_i = y_i = 0$.
This shows that points on the umbilic curve contained in $x_i = 0$ which 
have vanishing momentum $y_i = 0$ are indeed critical.
But what about other momenta? If the momentum $y_i$ is non-zero the
corresponding geodesic will leave the plane $x_i=0$. As soon as it is outside the
sub-ellipsoid ellipsoidal coordinates are regular, and thus the geodesic is
non-critical since the only critical points outside sub-ellipsoids are tori 
with fixed $\lambda_2$ with $\alpha_1 \le \lambda_2 \le \alpha_2$; 
but these tori have no point in common with 
the umbilic curve unless $\lambda_2 = \alpha_1$ or $\alpha_2$.

We have thus proved the well known result that the bifurcation diagram is obtained from 
collisions of roots of  the hyperelliptic curve $w^2 = -Q(z)A(z)$.
Subflows correspond to $Q(z)$ having a root that coincides with a root of $A(z)$. 
Hence the four sub-ellipsoids are given by the
lines $Q(\alpha_j) = 0$ in the image of the energy momentum map $(h, s_2, s_1)$.
For geodesic flows the energy can be fixed to $1/2$ without loss of generality, 
and thus the four lines in $(s_1, s_2)$ space are the straight lines 
$\alpha_j^2 + s_2 \alpha_j + s_1 = 0$, see figure~\ref{BifDiagGen}. 
These four lines intersect in six points $(s_1, s_2) = (\alpha_j \alpha_k, -\alpha_j - \alpha_k)$,
corresponding to periodic motion in the $jk$-plane.
The other curved line of the bifurcation diagram is given by double roots in $Q(z) = z (z - d)^2$
such that $(s_1, s_2) = (d^2, -2d)$ where $\alpha_1 \le d = \lambda_2 \le \alpha_2$ 
attaching tangentially to the straight lines of intermediate slopes
at the codimension two points $(\alpha_i^2, -2 \alpha_i)$, $i = 1,2$.
The four disjoint regions of regular values have 2 or 4 tori in their preimage. 

\begin{figure}
\centerline{\includegraphics[width=14cm]{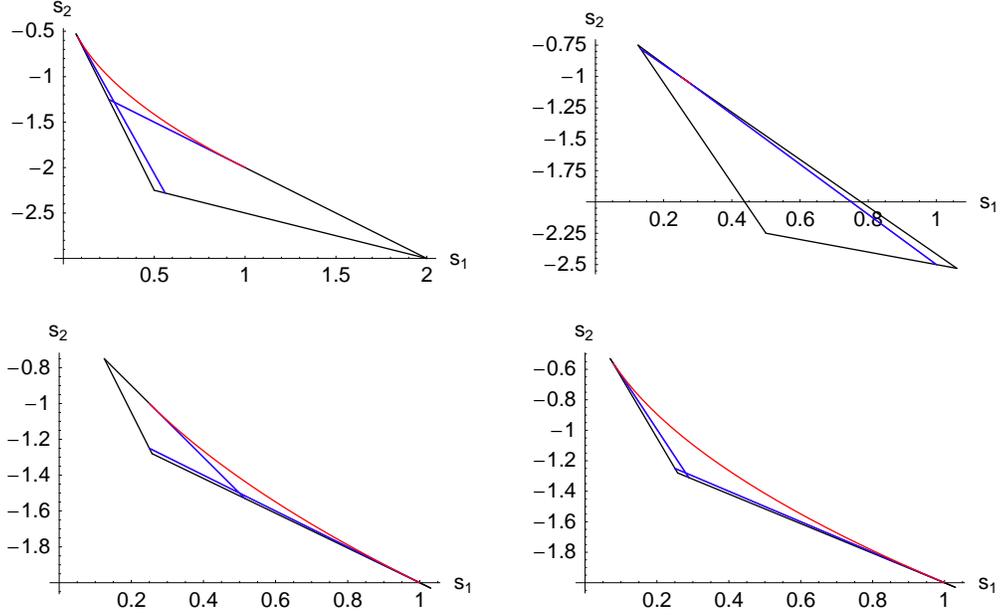}}
\caption{Bifurcation Diagram of almost degenerate Ellipsoids with 
$\alpha_i = (1/4, 1/4 + \eps, 1, 2)$, 
$(1/4, 1/2, 1/2 + \eps, 2)$, 
$(1/4, 1/2, 1, 1 + \eps)$, 
$(1/4, 1/4+\eps, 1, 1 + \eps)$, where $\eps = 0.03$.
} 
\label{DegenBifDiags}
\end{figure}

Finally it is important to establish the (non-)degeneracy and type~\cite{bolsinov}
of the singular points in phase space.
The gradient of $F_i$ vanishes in the plane $x_i =y_i = 0$ since each 
term is quadratic and contains the vanishing terms.
For non-degeneracy the spectrum of  the Jacobian 
matrix of the flow of $F_i$, which is denoted by $DX_{F_i}$ needs to be computed.
When restricted to the plane $x_i = y_i = 0$ the only nonzero entries
are in the $x_i$-$y_i$ sub-block. This sub-block reads
\begin{equation}
\label{DXFi}
\begin{pmatrix}
  -2 K_i(x,y) &  2(K_i(x, x) - 1) \\
  -2K_i(y, y) &  2K_i(x, y)  
\end{pmatrix} \qquad \text{where} \quad
K_i(x, y) = \sum_{k \not = i} \frac{x_k y_k}{\alpha_k - \alpha_i} \,.
\end{equation}
Notice that $K_i(y, y)$ never vanishes when $F_0$ and $F_3$ ($i=0,3$)
are considered. Otherwise not all terms of the same sign and $K_i(y,y)$
can vanish. 
Since this matrix is traceless the square of the eigenvalues is 
given by the negative determinant.
This condition needs to be evaluated on a point of the singular fibre in question.
Consider, say,  $F_1$. Then the point $x_0 = x_3 = 0$ is on all critical sets.  Then $y_2 = 0$ since this point is at a maximum of $x_2$ on the sub-ellipsoid $x_0 = 0$; moreover $x_2 = \pm \sqrt{\alpha_2}$.
At this point the diagonal terms vanish, and the eigenvalues vanish when 
$K_i(y , y)$ vanishes. For $F_1 = 0$ this occurs for the special momentum when 
$ y_0^2 = 2 h (a_0 - a_1)/(a_0 - a_3)$, and this is the point of tangency of the
line $F_1 =0$ with the curve of double roots in $Q$ in the bifurcation diagram. 
All other points are non-degenerate.
Similarly for $F_2$. This shows that there are two degenerate singularities in 
the geodesic flow on the non-degenerate $n$-dimensional ellipsoid when $n = 3$.
This is why the claim made in Theorem 2 of \cite{zung96}, that the geodesic flow on
the ellipsoid with distinct semi-axes is strongly nondegenerate, is not true.

All critical points corresponding to $F_0 = 0$ and $F_3 =0$ are non-degenerate
since $K_i(y,y)$ cannot vanish.
The type of the points (i.e. whether the eigenvalues are elliptic, hyperbolic,
or focus-focus) is elliptic on $F_0=0$ and $F_3=0$,
while it changes from elliptic to hyperbolic at the degenerate points 
on $F_1 = 0$ and $F_2=0$. This is where the straight line $F_1=0$
in figure~\ref{BifDiagGen} changes from being the outer boundary of the 
image of the momentum map to a line interior to the image; similarly for 
the line $F_2 = 0$.

The critical points with double roots $r_1 = r_2 = d$ are non-degenerate 
when they occur outside the umbilic curves. In that case the Jacobian matrix of 
the flow generated by the one degree of freedom system 
$\hat p_2^2 + Q(\lambda_2(\phi_2))$ is non-degenerate at $\lambda_2(\phi_2) = d$
since the 2nd derivative of $Q$ simply gives $2d$, which is non-zero,
and gives elliptic type.

The six corank two points given by the intersection of the lines $F_i = 0$
and $F_j= 0$ are non-degenerate. 
The above statements can now be specialised to the plane
$x_i = x_j = y_i = y_j = 0$. This is a one degree of freedom geodesic flow 
on the ellipse. The points on this orbit can be easily parametrized and   
then the eigenvalues become
$
\lambda_i^2 = 8 {h a_i}/{(-(a_k-a_i)(a_l-a_i))}
$ 
where $k,l$ are the other two indices distinct from $i,j$.
These eigenvalues are real or pure imaginary.
The linear combination $\alpha D X_{F_i} + \beta D X_{F_j}$
has eigenvalues $\alpha \lambda_i$ and $\beta \lambda_j$, 
which are different assuming $a_i \not = a_j$.
The combinations that occur  are elliptic-elliptic (indices 03, 01, 23), elliptic-hyperbolic (02, 13), and 
hyperbolic-hyperbolic (12). 

The topology of the bifurcation diagram is always that of figure~\ref{BifDiagGen} 
as long as all the semi-axes of the ellipsoid are distinct. 
As a first attempt to understand the degenerate cases in figure~\ref{DegenBifDiags} 
the bifurcation diagram is presented for four cases in which the 
semi-axes nearly coincide. From top left to bottom right the cases are close to 
equal smallest axes, equal middle axes, equal largest axes, and equal smallest
and largest axes, also denoted by 211, 121, 112, and 22. It appears as if
in the 121 and 22 case the image of the momentum map has only a single chamber,
and in the 121-case the image is a triangle. We will see that this
is not quite the correct answer. From the bifurcation diagrams it is clear, 
however, that the image of the symmetric subspace(s) which correspond to a 
number of collapsing lines needs to be analysed anew, while the results outside
this preimage can be taken over.

\section{Ellipsoid with equal middle axes}

Consider the geodesic flow on a three ellipsoid with equal middle axes $\alpha_1=\alpha_2$.  
The Casimirs $C_1$ and $C_2$ and the resulting Dirac bracket \eqref{eqn:dirac}  and the 
Hamiltonian are the same as before. However, the integrals $F_1$ and $F_2$ are not 
defined any more, but the singular terms cancel in the sum $G = F_1 + F_2$.
The other integrals $F_0$ and $F_3$ remain the same. 
The system is invariant under rotations in the $(x_1, x_2)$ plane and its cotangent lift, 
i.e.\ simultaneous rotation in the $(y_1, y_2)$ plane.  
This $SO(2)$ group action is
\begin{equation}
\Phi(x, y; \theta) = (\tilde x, \tilde y)
\end{equation}
where
\begin{equation}
\begin{aligned}
\tilde x &= (x_0, x_1\cos \theta-x_2\sin \theta, x_1\sin \theta+x_2\cos \theta, x_3) \\
\tilde y &= ( y_0, y_1\cos \theta-y_2\sin \theta, y_1\sin \theta+y_2\cos \theta, y_3) \, .
\end{aligned}
\end{equation}
The group action $\Phi$ is the flow generated by the angular momentum $J = x_1 y_2 - x_2 y_1$,
which is a global action variable since it generates the periodic flow $\Phi$.

\begin{theorem} {\em Liouville Integrability} \label{thm:ALInt}
The Geodesic flow on the ellipsoid with equal middle axes is Liouville integrable.  
Constants of motion are the energy $H=\frac {1} {2} \left(y_0^2+y_1^2+y_2^2+y_3^2\right)$, 
the angular momentum $J = x_1y_2 - x_2y_1$, 
and the third integral $G = F_1 + F_2$
\begin{equation}
G = y_1^2 + y_2^2 + \frac {\left(x_0y_1 - x_1y_0\right)^2} {\alpha_1 - \alpha_0} + \frac {\left(x_0y_2 - x_2y_0\right)^2} {\alpha_1 - \alpha_0} + \frac {\left(x_1y_3 - x_3y_1\right)^2} {\alpha_1 - \alpha_3} + \frac {\left(x_2y_3 - x_3y_2\right)^2} {\alpha_1 - \alpha_3}
\end{equation}
\end{theorem}

\begin{proof}
As in the generic case $2 H = F_0 + G + F_3$. So $G$ and $H$ commute because the $F_i$
commute in the generic case. In the limit $\alpha_2 \to \alpha_1$ the constant of motion
$(\alpha_1 - \alpha_2)F_1$ becomes $J^2$. Hence $J^2$ commutes with $H$ and $G$,
and therefore also $J$.

The constants of motion are not independent on the symplectic leaves of the Dirac bracket. 
Their relation is
\begin{equation} \label{FGrel}
   \frac{F_0}{\alpha_0} + \frac{G}{\alpha_1} - \frac{J^2}{\alpha_1^2} + \frac{F_3}{\alpha_3} = 0 \,,
\end{equation}
which is a straightforward limit of the generic relation $\sum F_i/\alpha_i = 0$
using $F_1/\alpha_1 + F_2/\alpha_2 = G/\alpha_2 - F_1(\alpha_1 - \alpha_2)/(\alpha_1 \alpha_2)$.

$H$, $J$, $G$ and the Casimirs $C_1$, $C_2$ are functionally independent almost everywhere on the level set $C_1 = C_2 =0$: They are polynomial and independent e.g. at $x=(\sqrt{\alpha_0}, 0, 0, 0)$, $y=(0, 1, 0, 0)$.
\end{proof}

The group action $\Phi$ has the invariants
\begin{equation}
\label{eqn:so2invts}
\pi_1 = x_1^2+x_2^2,\quad
\pi_2 = y_1^2+y_2^2,\quad
\pi_3 = x_1y_1 +x_2 y_2,\quad
\pi_4 = x_1 y_2 -x_2 y_1\,,
\end{equation}
related by $\pi_1\pi_2 -\pi_3^2 -\pi_4^2 =0$. The remaining variables $x_0,x_3,y_0,y_3$
are trivial invariants of $\Phi$. The fixed points of $\Phi$ have $x_1=x_2=y_1=y_2=0$.
When $J=\pi_4 = j \not = 0$ the fixed points are not in $J^{-1}(j)$ and the reduction by
the $SO(2)$ symmetry leads to a smooth reduced system on $J^{-1}(j)/SO(2)$:

\begin{lemma}  \label{thm:Regredn}
A set of reduced coordinates $(\xi_0,\xi_1,\xi_2,\eta_0,\eta_1,\eta_2)$ is defined on the reduced phase space $P_j=J^{-1}(j)/SO(2)$ by the formulae
\[
   \xi_0=x_0, \quad \xi_1=\sqrt{\pi_1}, \quad \xi_2 = x_3, \qquad
   \eta_0=y_0,\quad \eta_1 =  \frac{\pi_3}{\sqrt{\pi_1}}, \quad \eta_2 =y_3.
\]
The reduced coordinates satisfy the Dirac bracket in $\R^6[\xi,\eta]$, i.e.
\begin{equation}
\left\{\xi_k,\xi_l\right\} = 0, \qquad 
\left\{\xi_k,\eta_l\right\} = \delta_{kl} - \frac {\xi_k\xi_l} {D \alpha_k \alpha_l }, \qquad 
\left\{\eta_k,\eta_l\right\} = - \frac { \xi_k \eta_l - \xi_l \eta_k} {D \alpha_k \alpha_l} \,.
\nonumber
\end{equation}
The mapping $R:\R^8[x,y] \to \R^6[\xi,\eta]$ is a Poisson map and the reduced system has reduced Hamiltonian
\[
\hat H = \frac12 ( \eta_0^2 +\eta_1^2+\eta_2^2)+ \frac{j^2}{2\xi_1^2} 
\]
and additional integral
\[
\hat G = \eta_1^2 + \frac{(\xi_1\eta_0 - \xi_0 \eta_1)^2}{\alpha_1 - \alpha_0} 
                            + \frac{(\xi_1\eta_2 - \xi_2 \eta_1)^2}{\alpha_1 - \alpha_3}
      + \frac{j^2}{\xi_1^2} \left( 1 +  \frac{\xi_0^2}{\alpha_1 - \alpha_0} 
                                                        + \frac{\xi_3^2}{\alpha_1 - \alpha_3} \right)  \,.
\]
\end{lemma}

\begin{proof}
Define a set of coordinates on $\R^6[\xi,\eta]$ as shown.
The Poisson property of the map $R$, i.e. $\{ f \circ R, g \circ R\}_8 = \{f, g\}_6 \circ R$
follows from direct computation of the basic brackets, e.g.
$\{ \xi_1, \xi_2 \}_6 = \{ \sqrt{x_1^2+x_2^2}, x_3 \}_8 =0$, 
$\{ \xi_1, \eta_2 \}_6 = \{ \sqrt{x_1^2+x_2^2}, y_3 \}_8 = -(x_1^2x_3 +x_2^2x_3 )/(D \alpha_1 \alpha_3\sqrt{x_1^2+x_2^2}) = -\xi_1 \xi_2 /(D \alpha_1 \alpha_3)$, etc. The reduced bracket $\{,\}_6$ has
the Casimirs $\hat C_1 = \xi_0^2/\alpha_0 +\xi_1^2/\alpha_1 +\xi_3^2/\alpha_3-1$
and $\hat C_2 = \xi_0 \eta_0/\alpha_0 +\xi_1\eta_1/\alpha_1 +\xi_2 \eta_2/\alpha_2$.
The relation between the invariants becomes $\xi_1^2\pi_2 - \eta_1^2 \xi_1^2 -j^2 =0$ 
and elimination of $\pi_2$ from the Hamiltonian leads to the above result. 
Similarly the integral $G$ can first be written in terms of invariants $\pi_i$, $i=1,2,3$, 
and then the elimination of $\pi_i$ in addition using 
$\pi_1 = \xi_1^2$ and $\pi_3  = \eta_1 \xi_1$ gives the result.
\end{proof}

The reduced system is the ``geodesic flow" on the 2-dimensional ellipsoid
with semi-axes $\sqrt{\alpha_0}, \sqrt{\alpha_1}, \sqrt{\alpha_3}$ and an additional
effective potential $j^2/2\xi_1^2$. By definition $\xi_1 > 0$, so that the
reduced system for $|j|>0$ has only the open half of the ellipsoid as configuration space.
Since $|j|>0$ the plane $\xi_1=0$ is dynamically not accessible because
$\xi_1=0 \Rightarrow x_1=x_2=0 \Rightarrow j=0$.
Clearly the coordinates $x_0,x_3,y_0,y_3$ can serve as local 
coordinates on the half-ellipsoid, see the singular reduction below. 
Alternatively ellipsoidal coordinates on
the $\xi$-ellipsoid can be used to separate the variables.

To this end define a (singular) coordinate system on $\R^8[x_i,y_i]$ by
\begin{equation}
x_1 = \xi_1 \cos\theta, \qquad x_2 = \xi_1 \sin\theta
\end{equation}
where $\theta$ is the angle of rotation corresponding to the $SO(2)$ symmetry group action $\Phi$. 
The 2-ellipsoid embedded in $\mathbb{R}^3$ is defined by $\hat C_1=0$.
Coordinates $(\lambda_0, \lambda_1, \lambda_2)$ are then chosen as a confocal ellipsoidal 
coordinate system~\cite{moser80} in $\mathbb{R}^3[\xi]$, these being the roots $z$ of
\begin{equation}
\frac {\xi_0^2} {\alpha_0 - z} + \frac {\xi_1^2} {\alpha_1 - z} + \frac {\xi_2^2} {\alpha_3 - z} = 1.
\end{equation}
Constant $\lambda_0$ defines an ellipsoid, constant $\lambda_1$ a one-sheeted hyperboloid and constant $\lambda_2$ a two sheeted hyperboloid, where $\lambda_0 \le \alpha_0 \le \lambda_1 \le \alpha_1 \le \lambda_2 \le \alpha_3$.  Fixing $\lambda_0 = 0$ gives a set of generalised coordinates $(\lambda_1, \lambda_2, \theta)$ on the three ellipsoid with the middle two semi-axes equal.  The conjugate momenta are denoted by $(p_1, p_2, p_\theta)$, where  $p_\theta$ is the angular momentum $J$ found earlier.

\begin{lemma} \label{thm:Separation}
The Hamiltonian for the geodesic flow on the ellipsoid with equal middle axes in local symplectic coordinates $(\lambda_1, \lambda_2, \theta, p_1, p_2, p_\theta)$ reads
\begin{eqnarray}
H & = & - \frac {2(\alpha_0 - \lambda_1)(\alpha_1 - \lambda_1)(\alpha_3 - \lambda_1)} {\lambda_1(\lambda_2 - \lambda_1)}p_1^2 
- \frac {2(\alpha_0 - \lambda_2)(\alpha_1 - \lambda_2)(\alpha_3 - \lambda_2)} {\lambda_2(\lambda_1 - \lambda_2)} p_2^2 \nonumber \\ 
& + & \frac {(\alpha_0-\alpha_1)(\alpha_3-\alpha_1)} {2\alpha_1(\lambda_1 - \alpha_1)(\lambda_2 - \alpha_1)} p_\theta^2. 
\nonumber
\end{eqnarray}
The constants of motion are $p_\theta$ and $\tilde G_i$ % found by separating the variables as
\begin{equation}
\tilde G_i = \frac {2(\alpha_0 - \lambda_i)(\alpha_1 - \lambda_i)(\alpha_3 - \lambda_i)} {\lambda_i} p_i^2 - h\lambda_i - \frac {(\alpha_0-\alpha_1)(\alpha_3-\alpha_1)} {2\alpha_1(\lambda_i - \alpha_1)} p_\theta^2 \nonumber
\end{equation}
where $i =1,2$.  The integrals $G$ and $\tilde G_i$ are related by
\begin{equation}
\tilde G_1 + \tilde G_2 = \frac {(\alpha_1-\alpha_3)(\alpha_1-\alpha_0)G} {\alpha_1} -2\alpha_1h + \frac {\alpha_1^2 -\alpha_0\alpha_3} {\alpha_1^2}p_\theta^2. \nonumber
\end{equation}
\end{lemma}

\begin{proof}
The Hamiltonian in local coordinates is found after performing a cotangent lift of the new coordinates, and then expressing the original Hamiltonian in terms of those coordinates.  
The Hamiltonian is separated simply by multiplication with $\lambda_2 - \lambda_1$ and rearranging
to determine $\tilde G_1$ and $\tilde G_2$.
As a result of the separation the momenta $p_i$ conjugate to $\lambda_i$ can 
be expressed as
\begin{equation}
   p_i^2 = -\frac{\tilde Q( \lambda_i)}{4 A(\lambda_i)}
\end{equation}
with $\tilde Q$ being the analogue
of $Q$ in \eqref{PQU} given as
\begin{eqnarray} \label{tildeQ}
\frac{\tilde Q(z) }{z}
%& = & 2 h z^3 + 2 \left( \tilde g   - h \alpha_1\right) z^2 
% - \left( 2 \alpha_1 \tilde g + \frac {(\alpha_1-\alpha_0)(\alpha_3 - \alpha_1)} {\alpha_1}p_\theta^2\right)z \\
& = & 2h (\alpha_1 - z)^2 +  \\
&&  \frac{(\alpha_3 - \alpha_1)(\alpha_1-\alpha_0)}{\alpha_1}( g(\alpha_1-z) - j^2) +  \frac{\alpha_0 \alpha_3 - \alpha_1^2}{\alpha_1^2} j^2(\alpha_1 - z) \nonumber
\end{eqnarray}
The relation between the constant of motion $G$ and the separation constants $\tilde G_i$ is derived by substituting the expressions for the original coordinates in terms of the new local coordinates into $G$, rearranging and expressing in partial fractions.
\end{proof}
An analogue of the relation between the constants of motion for the generic 3-dimensional ellipsoid \eqref{eqn:FandS} is given by
\[
\frac{F_0}{z-\alpha_0}+ \frac{F_3}{z-\alpha_3}+ \frac{G}{z-\alpha_1}+ \frac{J^2}{(z-\alpha_1)^2}
= 
\frac{\tilde Q(z)} {A(z)} \,.
\]

The separating coordinate system is singular whenever $\lambda_i$ equals $\alpha_k$,
hence whenever $\xi_k = 0$.
When smooth elliptic coordinates $\phi_i$ are introduced the singularity at $x_1 = x_2 = 0$
remains, while those at $x_0=0$ and $x_3=0$ disappear.
The umbilical points on the reduced ellipsoid
$(x_0^2,x_3^2) = (\alpha_0(\alpha_1-\alpha_0), \alpha_3(\alpha_3-\alpha_1) )/(\alpha_3-\alpha_0)$
(coresponding to $\lambda_1 = \lambda_2 = \alpha_1$) are contained in the plane $\xi_1 = 0$.

Lemma \ref{thm:Separation} can be read as singular coordinates: for the full as for the reduced system.

\section{Singular Reduction}

The singular reduction for $j=0$ leads to a reduced system on a non-smooth manifold.
To understand its singularity let us consider as an aside the simple example of the $SO(2)$
 action $\Phi$ on the cotangent bundle $T^*\mathbb{R}^2$ with coordinates $(x_1,x_2,y_1,y_2)$. 
We can define the angular momentum, invariants and reduction map in exactly the same way as we did for the case of geodesic flow on the ellipsoid.  The reduction map $\pi$ gives a reduced phase space
given by the surface $\pi_1\pi_2-\pi_3^2 = j^2$ embedded in $\R^3[\pi_1,\pi_2,\pi_3]$.
This surface is a cone when $j=0$ and smooth otherwise. Considering the 
inequalities $\pi_1 \ge 0$,  $\pi_2\ge 0$ the reduced phase space for $j=0$
is half of a cone. An alternative description of this reduced phase space is
obtained by first restricting to any invariant subspace of the $x_1$-$x_2$-plane,
e.g. $x_2=y_2=0$. The $SO(2)$ action $\Phi$ has a residual $\Z_2$ action 
on this plane since $\Phi(\pi)(x_1,y_1) = (-x_1,-y_1)$. Therefore the singular 
reduced phase space $\{\pi_1 \pi_2 = \pi_3^2\} \subset \R^3$ can also be viewed as
$\R^2[x_1,y_1] / \Z_2$. This is e.g.\ the half-plane $x_1 \ge 0$ with the boundary $y_1=0$
identified with itself by $(x_1,0) \sim (-x_1,0)$, which again gives a cone.
The fixed point of the residual $\Z_2$ action $\Phi(\pi)$ is the origin $x_1=y_1=0$
and it is the singular point of the reduced phase space. Yet another representation of the 
same reduced phase space is given by classical polar coordinates $x_1 = r \cos\theta, x_2 = r \sin \theta$, so that the reduced space is the half plane $r,p_r$ with $r \ge 0$ and $p_r = (x_1 y_1 + x_2 y_2)/r$.
From the above we see that for $j=0$ the reduced space, however, is not a half-plane,
but a cone, because of the identification $p_r \sim -p_r$ along the line $r=0$.

The cone $\pi_1 \pi_2 = \pi_3^2$ can be diagonalised as a quadratic
 form by $\pi_1 = u + v$, $\pi_2 = u-v$, and $\pi_3 = w$. Then the cone
 is parametrised by $u = r$, $v = r\sin\phi$, and $w = r \cos\phi$. 
 In complex notation $z = r \exp i \phi$ the reduction map $\pi_1 = x^2$,
 $\pi_2 = y^2$, $\pi_3 = xy$ can then be written as $z = i (x - i y)^2/2$.
 Thus the mapping from $\R^2[x,y] \setminus (0,0)$ to the cone without tip is a 
 double cover. Therefore again the cone is equal to $\R^2 / \Z_2$.
 Moreover the Poisson-structure on the cone given by $\{\pi_1, \pi_2 \}_3 = 4 \pi_3$, 
 $\{\pi_1, \pi_3\}_3 = 2 \pi_1$, $\{\pi_2, \pi_3 \}_3 = -2 \pi_2$, 
 is mapped into the symplectic structure $\{\pi_1, \pi_2\}_3 = \{x^2, y^2 \}_2 = 4 x y = 4\pi_3$, 
 similarly for the other brackets. Moreover, the reduction map is invariant under the
 $\Z_2$ symmetry action, and thus the symplectic structure on the plane passes 
 down to the cone.  A similar argument is valid in the case of geodesic flow on the ellipsoid:

\begin{lemma} \label{thm:SingRedn}
The singular reduced phase space of the geodesic flow on the 3-ellipsoid with equal middle axes
and vanishing angular momentum $j=0$ is the phase space of the geodesic flow on the 2-ellipsoid reduced by the $\Z_2$ action 
$S(\xi_0,\xi_1,\xi_2,\eta_0,\eta_1,\eta_2) = (\xi_0,-\xi_1, \xi_2, \eta_0, -\eta_1,\eta_2)$.
Thus it is the geodesic flow on the 2-ellipsoid with a hard billiard wall inserted
in the $\xi_1=0$-plane.
\end{lemma}

\begin{proof}
The $SO(2)$ group action $\Phi$ does not act freely and so we have to use singular reduction to analyse the reduced phase spaces.  To do this we use invariant theory.  
The Casimirs for the system, expressed in terms of the invariants \eqref{eqn:so2invts}, are
\begin{equation}
\frac {x_0^2} {\alpha_0} + \frac {\pi_1} {\alpha_1} + \frac {x_3^2} {\alpha_3} = 1, \qquad \frac {x_0y_0} {\alpha_0} + \frac {\pi_3} {\alpha_1} + \frac {x_3y_3} {\alpha_3} = 0.
\label{eqn:Casimirs}
\end{equation}
Note that these equations are linear in the invariants.  
The Hamiltonian may be expressed in terms of the invariants as
\begin{equation}
H = \frac {1} {2} \left(y_0^2 + \pi_2 + y_3^2\right).
\label{eqn:engeqn}
\end{equation}
The reduced phase space is a subset of $\R^7[x_0, y_0, x_3, y_3, \pi_1, \pi_2, \pi_3]$.  
It is defined by the two Casimirs (\ref{eqn:Casimirs}), 
the relation between the invariants $\pi_1\pi_2 - \pi_3^2=j^2$ 
and the inequalities $\pi_1\ge0$, $\pi_2\ge0$.  
It carries the induced Poisson-structure.
To describe this subset we first of all eliminate $\pi_1$ and $\pi_3$ using (\ref{eqn:Casimirs}) to get
\begin{equation} \label{eqn:casi5}
\alpha_1\left(1 - \frac {x_0^2} {\alpha_0} - \frac {x_3^2} {\alpha_3}\right)\pi_2 -\alpha_1^2\left(\frac {x_0y_0} {\alpha_0} + \frac {x_3y_3} {\alpha_3}\right)^2 = j^2 \, .
% \label{eqn:redph}
\end{equation}
This is a single equation in $\R^5[x_0,x_3,y_0,y_3,\pi_2]$ which defines a four dimensional object.  Equating the gradient to zero, we find that the equation defines a smooth four dimensional manifold except when $j=0$.  This is the reduced phase space $P_j =J^{-1}(j)/SO(2)$.
When $j=0$ there is a singularity for $\pi_1= \pi_2=\pi_3=0$. The singular points
are given by the phase space of the geodesic flow on the ellipse in the 03-plane,
which is a cylinder. Thus the singular set of the reduced phase space is itself
a symplectic manifold. This symplectic manifold is invariant under the flow of the
reduced equations, but it is not fixed under it.

The geodesic flow on the 2-ellipsoid reduced by the $\Z_2$ action $S$ is the
billiard. Take as a fundamental region the half-ellipsoid with $\xi_1 \ge 0$.
Then the action of $S$ on the boundary $\xi_1=0$ simply flips the sign
of $\eta_1$, which can be viewed as the reflection on the plane $\xi_1=0$
with the rule ``angle of incidence equals angle of reflection".
Moreover, points $\xi_1=\eta_1=0$ are fixed under $S$ and correspond to
orbits that are sliding along (or in) the billiard boundary $\xi_1=0$.

To establish the correspondence between the reduced space and the billiard
consider the slice $x_2=y_2=0$ through full phase space. This is a geodesic
subflow, which is the geodesic flow on the 2-ellipsoid with semi-axes
$\sqrt{\alpha_0}$, $\sqrt{\alpha_1}$, $\sqrt{\alpha_3}$, as described in
lemma~\ref{thm:Regredn}.
Any  motion with $j=0$ can be reduced to a motion in this plane by some rotation
$\Phi(\theta)$ with constant $\theta$. So locally the flow on this 2-ellipsoid
is the reduced system. Globally, however, we still have the residual $\Z_2$ action $S$, given by
$\Phi(\pi)$, to reduce by.

The reduced bracket in $\R^5$ is the original Dirac bracket \eqref{eqn:dirac} 
between  $x_0, x_3, y_0, y_3$
with the additional non-zero brackets with $\pi_2$ given by 
$\{ y_i , \pi_2 \} = 2 x_i \pi_2 / ( \alpha_i \alpha_1 D)$. 
As in the trivial example above we now show that the 
mapping from the 2-ellipsoid minus a cylinder to the reduced phase space 
minus the singular set is a Poisson map.
When the singular set is removed, \eqref{eqn:casi5} can be solved for $\pi_2$. 
After elimination of $\pi_2$ only the original Dirac bracket between $x_0, x_3, y_0, y_3$ remains.
For the description of the geodesic flow on the 2-ellipsoid we use the the variables
$\xi_0, \xi_1, \xi_2$ and momenta $\eta_0, \eta_1, \eta_2$ which satisfy the Dirac bracket 
\eqref{eqn:dirac}, without implying that they are obtained by regular reduction as
in lemma~\ref{thm:Regredn}.
The mapping from the symplectic submanifold of $\R^6[\xi_0,\xi_1,\xi_2, \eta_0, \eta_1, \eta_2]$ 
obtained by fixing the Casimirs to $\R^4[x_0,x_3, y_0, y_3]$ (without any Casimirs)
is simply the projection $x_0 = \xi_0, x_3 = \xi_2, y_0 = \eta_0, y_3 = \eta_2$, which preserves
the Dirac bracket.
However, this mapping is $2:1$ since from the Casimirs only $\xi_1^2$ can
be recovered, but not its sign.

The meaning of this construction is very simple.
Because of the inequality $\pi_1 > 0$ the variables $x_0, x_3$ are restricted to 
the interior of the  ellipse $\pi_1 = 0$. These variables are good local coordinates 
on the reduced phase space after the singularity (at $\pi_1 = 0$) is removed. 
From the point of view of the 2-ellipsoid this amounts to choosing local coordinates
in configuration space as the projection of the point onto the $\xi_1 = 0$ plane.
\end{proof}

As a result the regular reduction described in lemma~\ref{thm:Regredn} gives 
the right description even in the singular case, when the additional discrete quotient
by the $\Z_2$ action $S$ is included in the picture. In the regular case $j\not = 0$ 
the two halves of the ellipsoid that are identified by $S$ are dynamically disconnected, 
because of the effective potential $j^2/2\xi_1^2$. However, when $j=0$ the $\Z_2$ action
is less trivial because its fixed set is now accessible to the dynamics, 
and this fact is crucial in order to obtain the correct singular fibres in the next section.

\section{The Liouville foliation}

We now wish to investigate the topology of the invariant level sets obtained by fixing the constants of motion.  The energy momentum map is ${\cal EM} = (H, J, G) : M \rightarrow \mathbb{R}^3$.  Since $H$ for a geodesic flow is homogeneous in the momenta we can fix the energy to, say, $h$.

\begin{theorem} \label{thm:EM}
The image of the energy momentum map ${\cal EM}$ for constant energy $H=h$ is the region in $\mathbb{R}^2$ bounded by the quadratic curves (see figure~\ref{fig:BiDi121})
\begin{equation}
g = \frac {2\alpha_1} {\alpha_1-\alpha_3} h - \frac {\alpha_3} {\alpha_1(\alpha_1-\alpha_3)} j^2, \qquad
g = \frac {2\alpha_1} {\alpha_1-\alpha_0} h - \frac {\alpha_0} {\alpha_1(\alpha_1-\alpha_0)} j^2.
\label{eqn:bdrycurves}
\end{equation}
Singular values of the energy momentum map are the boundary curves (elliptic), 
their intersections (elliptic-elliptic), and an isolated singularity at the origin $(j,g)=(0,0)$ 
of focus-focus type.
\end{theorem}

\begin{proof}
As in the generic case critical points can occur on sub-ellipsoids.
On $x_0 = y_0 = 0$ the integral $F_0 = 0$ and $\nabla F_0 = 0$,
similarly for $x_3 = y_3 = 0$ and $F_3$. In both cases the corresponding
sub-ellipsoids are ellipsoids of revolution. 
The image of the critical points
with $x_0 = y_0 = 0$ is found using the relation \eqref{FGrel}
to eliminate $F_3$ in $2H = F_0 + G + F_3$, which gives
\[
    2H = G - \alpha_3\left( \frac{G}{\alpha_1}  - \frac{J^2}{\alpha_1^2} \right),
\]
and hence the first curve of critical values. 
A similar computation for critical points with $x_3 = y_3 = 0$
gives the other curve. 

These points are non-degenerate because
the Jacobian of the flow generated by $F_0$ restricted to the critical points 
$x_0 = y_0 = 0$ is given by \eqref{DXFi}. Evaluating this on the point
$x_0 = x_1 =  x_3 = 0$, $y_0 = y_2 = 0$ shows that the eigenvalues of this matrix
never vanish and are of elliptic type, and hence these critical points are non-degenerate.
Similar arguments apply to the points $x_3 = y_3 = 0$.
The two corank two points given by the intersection of the two curves 
are also non-degenerate, because the non-zero $2\times 2$ blocks
of the Jacobians are distinct, so that $\mu DX_{F_0} + \nu DX_{F_3}$
spans the Cartan subalgebra; the 4 eigenvalues (for any point on the critical circles
given by $x_0 = x_3 = y_0 = y_3 = 0$) are
$\pm2  \imag \mu  \sqrt{2 \alpha_0 h}/(\alpha_1 - \alpha_0)$ and 
$\pm2 \imag \nu \sqrt{2\alpha_3h }/(\alpha_3 - \alpha_1)$.
This orbit is a relative equilibrium, i.e. a circle in the $x_1$-$x_2$ plane.
The eigenvalues of $DX_{F_0}$ and $DX_{F_3}$ are elliptic,
so at their intersection an orbit of elliptic-elliptic type is found.

Since for $x_i = y_i = 0$ the integrals $F_i = 0$ and
also its gradient vanishes, $G = F_1 + F_2$ and
its gradient clearly vanishes when 
$x_1 = x_2 = y_1 = y_2 = 0$.
Considering the Casimirs the solutions set of $x_1 = x_2 = y_1 = y_2 = 0$ 
is a geodesic flow on the ellipse in the $x_0$-$x_3$ plane. Fixing the
energy two critical circles are obtained.
On these critical points also $J =0$ so that the origin in the image $(J, G) = (0,0)$ 
is a critical value.  

Moreover $\nabla J = 0$ as well,
and the Jacobian of $X_J$ has eigenvalues $\pm \imag$, 
since  its flow $\Phi$ is a rotation. 
Finally $\mu DX_{G} + \nu DX_{J}$ spans the Cartan subalgebra; 
the 4 eigenvalues (for any point on the critical circles) are
$ \pm \mu \sqrt{ 8 \alpha_1 h / (\alpha_1-\alpha_0)(\alpha_3-\alpha_1)} \pm \imag \nu$.
These eigenvalues form a complex quadruplet, and hence the isolated critical point
at the origin is of focus-focus type.

This establishes the existence, non-degeneracy, and type of all critical points.
The bifurcation diagram is shown in figure~\ref{fig:BiDi121} with an isolated singularity at 
the origin $(j, g) = (0, 0)$.
The remaining part of the proof shows that there are no other critical points.

First of all the ellipsoidal coordinates from lemma~\ref{thm:Separation}
are used to establish that almost all other points are non-singular. 
These coordinates are non-singular outside any sub-ellipsoid $\xi_i = 0$. 
To find critical points in the region of phase space 
with $\xi$ coordinates such that all $\xi_i \not = 0$
it is enough to compute the rank of the matrix $D(\tilde G_1, \tilde G_2, p_\theta)$,
see lemma.~\ref{thm:Separation}. 
Since the variables are separated this implies $p_s = 0$ 
% ($\lambda_s = \alpha_k$ is forbidden by $\xi_i \not 0)
and 
$h 2\alpha_1(\lambda_s - \alpha_1)^2 = -(\alpha_1-\alpha_0)(\alpha_3-\alpha_1) p_\theta^2$
which is impossible. Thus critical points are contained in the coordinate singularities
$\xi_i=0$.

It remains to check the pre-images in full phase space 
of the sub-ellipsoids $\xi_i = 0$ where the ellipsoidal coordinates are not defined.
When $\eta_i = 0$ in addition to $\xi_i=0$ then the point is critical, see above.
Thus we need to show that all points with $\xi_i = 0$ but $\eta_i \not = 0$ are non-singular.
The ellipsoid $x_0=0$ is a totally geodesic submanifold, i.e. when $y_0 = 0$ every orbit
stays inside $x_0 = 0$. Conversely, when $y_0 \not = 0$ the orbit must leave the 
sub-ellipsoid $x_0 = 0$. Similar for $x_3=0$.
In general an orbit with $x$ in some sub-ellipsoid(s)
but $y$ not tangent to these sub-ellipsoids will leave them, and therefore
will have all $x_k \not = 0$. But there ellipsoidal coordinates are regular, 
and therefore the original point is non-critical, since geodesic motion
preserves non-criticality.  When $\xi_1 = 0$ and hence $x_1 = x_2 = 0$ the condition $\eta_1 = 0$ is always satisfied by definition, but it does not specify $y_1$ and $y_2$. 
But the previous argument applies again: if $(y_1, y_2) \not = (0,0)$ then 
the geodesic will leave $x_1 = x_2 = 0$, and therefore the original point is not critical.

\begin{figure}
\centerline{\includegraphics[width=8cm]{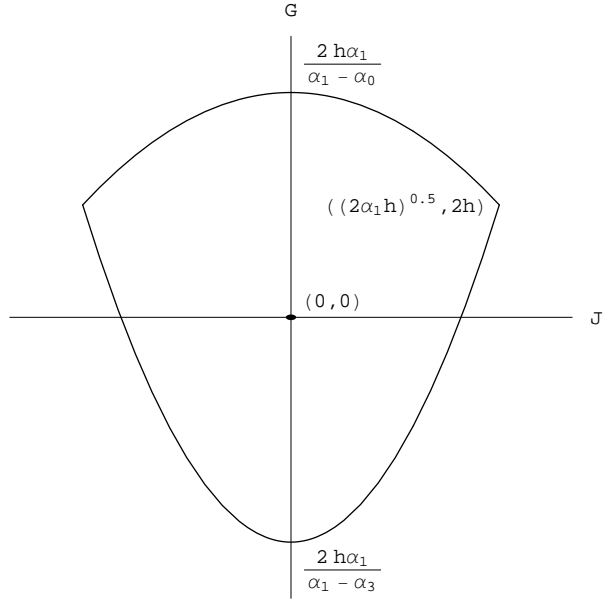}}
\caption{Bifurcation Diagram for the ellipsoid with axis $\alpha_0 < \alpha_1 = \alpha_2 < \alpha_3$.}
\label{fig:BiDi121}
\end{figure}

\end{proof}

The bifurcation diagram figure~\ref{fig:BiDi121} can be considered as the square root of the
diagram in figure~\ref{DegenBifDiags} top right. The line $J=0$ is where
the lines $F_1 = 0$ and $F_2 = 0$ collapse. However, in the limit coming 
from the generic case the whole line would appear to be critical,
since it is on the boundary of the image of the energy-momentum map.
Recall that the limit of $F_2(\alpha_2 - \alpha_1)$ (or $F_1(\alpha_1 - \alpha_2)$)
equals $J^2$, but not $J$. Obviously $J^2$ is singular when $J=0$, but 
$J$ itself is not. Thus the square root of figure~\ref{DegenBifDiags} top right
gives figure~\ref{fig:BiDi121}. Upon this transition most of the critical 
points along the lines $F_1 = F_2 = 0$ disappear, except for the 
isolated critical point, which can be thought of as the remainder of 
the intersection of the two lines and the critical curve corresponding to 
the umbillic line $\lambda_1 = \lambda_2$ (in the generic system).
Moreover the two corank 2 points at the corners of the triangle in 
figure~\ref{DegenBifDiags} top right only have corank 1 after passing from $J^2$
to the ``better'' constant of motion $J$. Finally, the multiplicity of the regular $T^3$
changes from 2 to 1 for every regular point in the image.
%Except for the multiplicity the preimages of values outside $J=0$ don't change in 
%the transition, but those with $J=0$ have to be computed anew.

The fibre of a regular value in the image of ${\cal EM}$ is a $T^3$ by the Liouville-Arnold theorem.  
We now wish to find the fibres of the energy momentum map at the singular values in the bifurcation diagram,
in particular at the isolated critical value.

\begin{theorem} \label{thm:SingFib}
The singular fibres over the boundary curves of the image of the energy momentum map at constant energy, with the exception of their intersections, are two tori $T^2$.  At each intersection point of the boundary curves the singular fibre is $S^1$.  The singular fibre over the isolated singularity at the origin is the direct product of $S^1$ and a doubly pinched torus $T^2$.
\end{theorem}

\begin{proof}
At the boundary of the image all singularities are of elliptic type, and hence 
the singular fibre is $T^{3-r}$ where $r$ is the corank of the singularity;
$r=1$ on the upper and lower curve and $r=2$ at their intersection points.

The upper boundary with $F_0=0$ consists of all orbits in the geodesic flow on the ellipsoid
of revolution defined by $x_0=y_0=0$. Reduction maps each $T^2$ of this system to a 
relative periodic orbit. The isloated periodic orbit in the $12$-plane of the geodesic flow 
on the ellipsoid of revolution corresponds to the extremal points with $J=\pm\sqrt{2\alpha_1 h}$.
Reduction maps this realtive equilibrium to the fixed point $\xi = (0, \sqrt{\alpha_1}, 0)$ 
on the middle-axis of the reduced ellipsoid.
A similar statement holds for the lower boundary $F_3=0$. The isolated periodic orbit in the 
$12$ plane is common to both ellipsoids of revolution.

\begin{figure}
\centerline{\includegraphics[width=6cm]{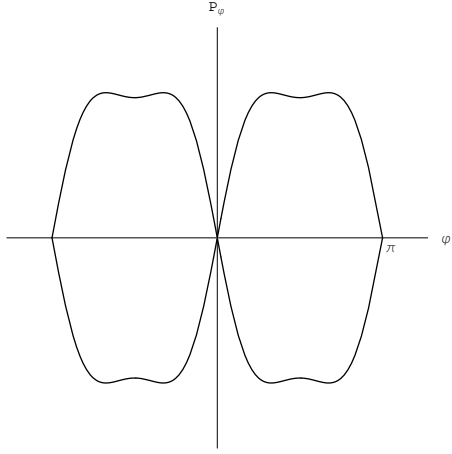}}
\caption{The intersection of the preimage of the isolated singular point $(j, g)=(0, 0)$ with the Poincare section $x_3=0$ in reduced phase space. The separatrix is of type C2 before quotient by $\Z_2$.}
\label{fig:fig2}
\end{figure}

The singular fibre for the isolated critical value at the origin of the bifurcation diagram  
contains two circles of critical points in the $03$-plane, see above.  
The critical points are non-degenerate and of focus-focus type. 
From the general theory \cite{zung96} it follows that it is an almost direct product of a
pinched torus multiplied by $S^1$.
Since there are two circles of critical points in the singular fibre the number of
pinches is 2. 

The fibre over the isolated singularity is complicated as it is not of elliptic type and thus contains 
the non-critical points of the separatrix in addition to the two circles of singular points.  
By lemma~\ref{thm:SingRedn} the reduced system for $j=0$ is the geodesic flow on the 2-ellipsoid 
quotient by the $\Z_2$ action $S$. Ignoring the quotient the reduced singular fibre consists 
of the unstable isolated periodic orbits in the plane $\xi_1=0$ and their separatrix.
In \cite{bolsinov} it was shown that the topology of this singular fibre is 
$C_2 \times S^1$, where $C_2$ stands for two circles intersecting in two points.
This can be seen from the Poincar\'e section $\xi_2=0$. Since $\xi_2 = \eta_2 =0$ 
is an invariant subflow the boundary of the section with $\eta_2 \ge 0$ is an 
invariant set and it is the only place where the flow is not transverse to the section.
In configuration space the section condition is the ellipse in the $01$-plane,
and it can be parametrised by an angle $\phi$ by
$(\xi_0,\xi_1) = ( \sqrt{\alpha_0} \cos\phi, \sqrt{\alpha_1} \sin\phi )$.
The momentum $p_\phi$ conjugate to $\phi$ then gives the momenta as
$(\eta_0,\eta_1) = ( \sqrt{\alpha_0} \sin\phi, -\sqrt{\alpha_1} \cos\phi ) p_\phi/d $
where $d = \alpha_0\sin^2(\phi) + \alpha_1\cos^2(\phi)$.
The reduced Hamiltonian can be solve for $\eta_2$ on the section and thus
the integral $G$ can be written as a function of $(\phi, p_\phi)$ on the section:
%\begin{equation}
%   g\frac{\alpha_1-\alpha_3}{\alpha_1} + 2 h \sin^2\phi  = p_\phi^2\left(d \sin^2\phi 
%   + \frac{\alpha_1-\alpha_3}{2(\alpha_0 - \alpha_1)} 
%     \left( \alpha_0 + \alpha_1 + (\alpha_1-\alpha_0)\cos 2\phi \right)
%   \right) \,.
%\end{equation}
\begin{equation}
   2 h \frac{\sin^2\phi}{\alpha_3-\alpha_1}  =    -g\frac{1}{\alpha_1} +
   \frac {p_\phi^2} {d^2} \left( \frac{d \sin^2\phi }{\alpha_3-\alpha_1}
   + \frac12\left (\frac{\alpha_1 + \alpha_0} {\alpha_1 - \alpha_0}  +  \cos 2\phi\right)
   \right) \,.
\end{equation}
The singular fibre is $g=0$ ($j=0$ was already used in the singular reduction),
and it defines two circles winding around the $(\phi,p_\phi)$ cyclinder 
intersecting at the points $(0,0)$ and $(\pi, 0)$, which are critical points of $g$. 
This is the ``atom" $C_2$~\cite{bolsinov}. 
Now the quotient with respect to $S$ has to be performed. In the new coordinates
the action of $S$ is $(\phi, p_\phi) \to (-\phi, -p_\phi)$, which fixes $\xi_0$ 
and $\eta_0$, but flips the sign of $\xi_1$ and $\eta_1$. This action has two fixed
points $(0,0)$, and $(\pi, 0)$, so that the fixed points of $S$ coincide with
the critical points of $g$. Reduction of the cylinder by the $\Z_2$ action 
gives the ``canoe" \cite{cushman97}, with two singular points. The two singular points are connected
by two half-circles. This is $C_2/\Z_2$, where $\Z_2$ acts by reflection such that the 
intersection points are fixed.

Since the reduced flow is transverse to the section on the singular fibre 
the complete reduced singular fibre is $(C_2 / \Z_2) \times S^1$.
The singular fibre in full phase space is found by letting $\Phi$ act
on the preimage of this set under the reduction map. 
Since the singular circles are fixed under $\Phi$ they will
remain singular circles, while every other point will be multiplied by $S^1$.
Exchanging the order of the operations, first acing with $\Phi$ on the 
preimage of $C_2 / \Z_2$ gives a double pinched torus, which is then
multiplied by $S^1$. This $S^1$ action also has a generator, which is a 
second global smooth action, see below.

\end{proof}

Following the approach suggested by Nguyen Tien Zung in \cite{zung96a}, one can reformulate the last statement of Theorem \ref{thm:SingFib} by saying that the singularity corresponding to the isolated singular point is the direct product of the standard 4-dimensional focus-focus singularity with 2 pinches and a non-singular system with 1 degree of freedom.  This is a kind of ``almost direct product" decomposition which can be found for any non-degenerate singularity (see \cite{zung96a}).  The fact that in our case the product is ``direct" seems to be a general property of focus type singularities (Nguyen Tien Zung, private communication).

\section{Actions and Monodromy}

We found that the equations for geodesic flow are Liouville-Arnold integrable, and so the fibre over a regular point is a $T^3$.  Let $C_1$, $C_2$, $C_3$ be a basis of cycles on this torus.  Due to the fact that the variables can be separated as in lemma~\ref{thm:Separation}, 
natural cycles are
\begin{equation}
C_1 : d\lambda_1=d\lambda_2=0, C_2 : d\theta=d\lambda_2=0, C_3 : d\theta=d\lambda_1=0 \,.
\end{equation}
The adjective `natural' is used in the technical sense of \cite{dullin02}, and simply means
to consider the obvious choice given by coordinate lines of the separating coordinate systems.
Natural though this may be, it will turn out that the corresponding actions are only continuous,
but not smooth.
The  corresponding natural action variables  are given by
\begin{equation}
\label{eqn:acts}
I_1 = \frac {1} {2\pi} \oint_{C_1} p_\theta d\theta = p_\theta, \qquad I_2 = \frac {m_2} {2\pi} \oint_{C_2} p_1 d\lambda_1, \qquad I_3 = \frac {m_3} {2\pi} \oint_{C_3} p_2 d\lambda_2.
\end{equation}
The first action is just the angular momentum.  The momentum $p_i$ in the second and third actions 
is given in lemma~\ref{thm:Separation}, together with the polynomial $\tilde Q$ in  \eqref{tildeQ}.
The integer multipliers $m_2=2$ and $m_3=2$ arise due to the way in which the ellipsoidal 
coordinates are set up over the eight octants in $\mathbb{R}^3$, see~\cite{dullin01}.  When $I_2$ (respectively $I_3$) is evaluated on the upper (respectively lower) boundary of the bifurcation diagram (Figure \ref{fig:BiDi121}) the natural actions for the geodesic flow on the ellipsoid of revolution are found, see appendix.

\begin{figure}
\centerline{\includegraphics[width=8cm]{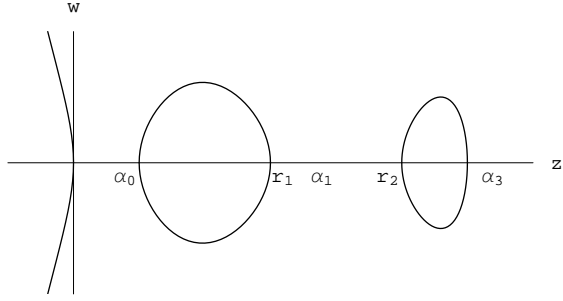}}
\caption{Real part of curve $w^2=-z(z-\alpha_0)(z-r_1)(z-r_2)(z-\alpha_3)$ showing cycles.}
\label{fig:fig3}
\end{figure}

The polynomial $\tilde Q$ can be factored as $\tilde Q(z)=z(z-r_1)(z-r_2)$ where,
in order to have real momenta $p_1$, $p_2$,
\begin{equation}
\alpha_0 \le r_1 \le \alpha_1 \le r_2 \le \alpha_3 \,.
\end{equation}
The integrals \eqref{eqn:acts} are calculated on the hyperelliptic curve given by 
\begin{equation}
w^2 =  - \frac{ A(z)\tilde Q(z) }{ (z-\alpha_1)^2}.
% \quad \text{ where } \quad
 % \tilde A(z) = (\alpha_0 - z)(\alpha_1 - z)(\alpha_3 - z) \,.
\end{equation}
The genus of this curve is one less than for a non-degenerate ellipsoid
because the pole in $A$ can be divided out.
This is plotted in figure \ref{fig:fig3} and the cycles can be seen.  
Note that the part of the curve in the negative $z$ range does not correspond to any real motion.
Writing out the actions in full we have proved the following theorem

\begin{lemma} \label{thm:actions}
The actions of the geodesic flow on the three dimensional ellipsoid with equal middle axes are given by $I_1 = p_\theta$,
\begin{equation}
I_2 = \frac {1} { \pi } \oint_{C_2} \frac {\tilde Q}{2 (\alpha_1 - z) w} \, \dee z 
%\end{equation}
\quad \text{ and } \quad
%\begin{equation}
I_3 = \frac {1} { \pi } \oint_{C_3} \frac {\tilde Q}{2 (z-\alpha_1) w}  \, \dee z \,.
\end{equation}
\end{lemma}

\begin{figure}
\centerline{\includegraphics[width=6cm]{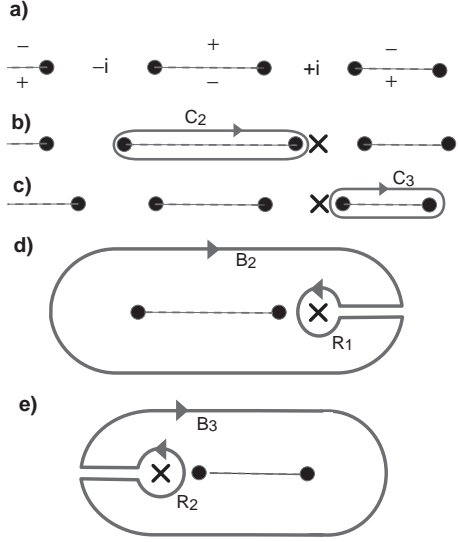}}
\caption{Complex plane $\mathbb{C}(s)$ and choice of branch cuts (a), integration paths for $p_\theta$ for $r_1\rightarrow \alpha_1$ (b) and $r_2 \rightarrow \alpha_1$ (c).  (d) and (e) show decomposition of $C_2$ for case (b) and $C_3$ for case (c).}
\label{fig:fig4}
\end{figure}

Note that the constants of motion $p_\theta$, $h$ and $g$ are implicit in the definition of $\tilde Q(z)$ given in \eqref{tildeQ}.  The integrand has a simple pole at $\alpha_1$ and branch points at  $\alpha_0$, $\alpha_3$, and at $r_1$ and $r_2$ for $p_\theta \ne 0$.  The integrals are hyper-elliptic of genus 2 and third kind.  The three natural actions $I_1$, $I_2$ and $I_3$ are functions of $p_\theta$, $g$ and $h$.  However, we will show that $I_2$ and $I_3$ are not differentiable at $p_\theta=0$.  The derivative of these actions with respect to $p_\theta$ is given by
\begin{equation}
\frac {\partial I_i} {\partial p_\theta} = 
-\frac {(\alpha_3-\alpha_1)(\alpha_1-\alpha_0) p_\theta} {2 \pi\alpha_1} 
\oint_{C_i} \frac{z}{w(\alpha_1 - z)} \, \dee z
\end{equation}
Figure \ref{fig:fig4} indicates the poles, branch points and integration paths.  Note that as $p_\theta \rightarrow 0$, then we find that one of the branch points $r_i$ tends to the pole at $\alpha_1$.
\begin{equation}
g<0 \Rightarrow \lim_{p_\theta \rightarrow 0} r_2 = \alpha_1, \qquad g>0 \Rightarrow \lim_{p_\theta \rightarrow 0} r_1 = \alpha_1.
\end{equation}

We follow the approach of Cushman \cite{cushman97} and Dullin~\cite{dullin02} and deform the integration path so that the integral may be split up into three or two separate integrals depending upon which case is being considered.  The integral around the branch points $C_i$ is expanded into loops $B_i$ around the poles and then the contributions from the poles $R_1, R_2$ are subtracted.
\begin{equation}
\oint_{C_2} = \oint_{B_2} + \oint_{R_1}, \qquad \oint_{C_3} = \oint_{B_3} + \oint_{R_2}
\end{equation}
Now evaluating the residue of the integrand at the simple pole, we have
\begin{equation}
\res_{z=\alpha_1} \frac {z}{(z-\alpha_1) w} \, \dee z = \frac {\alpha_1 } {2(\alpha_3-\alpha_1)(\alpha_1-\alpha_0)i|p_\theta|}
\end{equation}
We then have for $g<0$,
\begin{equation}
\lim_{p_\theta \rightarrow 0} \frac {\partial I_2} {p_\theta} = 0, \qquad \lim_{p_\theta \rightarrow 0^+} \frac {\partial I_3} {\partial p_\theta} =  -1, \qquad \lim_{p_\theta \rightarrow 0^-} \frac {\partial I_3} {\partial p_\theta} = 1
\end{equation}
and for $g>0$,
\begin{equation}
 \lim_{p_\theta \rightarrow 0^+} \frac {\partial I_2} {\partial p_\theta} = -1, \qquad \lim_{p_\theta \rightarrow 0^-} \frac {\partial I_2} {\partial p_\theta} = 1, \qquad \lim_{p_\theta \rightarrow 0} \frac {\partial I_3} {p_\theta} = 0.
\end{equation}
In other words we have
\begin{equation}
\lim_{p_\theta \rightarrow 0^-} \frac {\partial I_i} {\partial p_\theta} \neq \lim_{p_\theta \rightarrow 0^+} \frac {\partial I_i} {\partial p_\theta}, \qquad i=2,3 \nonumber
\end{equation}
Since $I_2$ and $I_3$ are even functions of $p_\theta$, differentiability at $0$ would imply that the derivative is zero at $p_\theta=0$.  But there is a discontinuity in the derivative here and so the natural actions are continuous but not differentiable at $p_\theta=0$.  By changing the basis of cycles locally smooth actions can be found, but they are then globally multi-valued.  We describe this in the following manner.
\\
\\
Let the natural actions for positive $p_\theta$ be represented by $I_+=(I_1,I_2,I_3)^t$, and those for negative $p_\theta$ by $I_-$.  $I_1$ is odd and $I_2$, $I_3$ are even, hence
\begin{equation}
I_-(-p_\theta) = SI_+(p_\theta)
\end{equation}
where $S = {\rm diag}(-1, 1, 1)$.
%\begin{equation}
%S = \left(
%\begin{array}{ccc}
%-1 & 0 & 0 \\
%0 & 1 & 0 \\
%0 & 0 & 1
%\end{array}
%\right)
%\end{equation}
We then define unimodular matrices $M_i$ such that $I_+$ and $M_iI_-$ join smoothly at $p_\theta=0$.  This smoothes the actions locally.  For continuity on the line $p_\theta=0$ it is necessary that below the isolated critical point
\begin{equation}
I_+ = M_1I_- = M_1SI_+ = M_1I_+, \qquad g>0
\end{equation}
and above
\begin{equation}
I_+ = M_2I_- = M_2SI_+ = M_2I_+, \qquad g<0
\end{equation}
Therefore $(0, I_2, I_3)^t$ is an eigenvector of $M_i$ with eigenvalue $1$~\cite{dullin04}.  The eigenvector equations show that $M_i$ must have the form
\begin{equation}
M_i = \left(
\begin{array}{ccc}
\delta_i & 0 & 0 \\
\kappa_i & 1 & 0 \\
\beta_i & 0 & 1
\end{array}
\right)
\end{equation}
and since $M_i \in SL(3,\mathbb{Z})$ as it is unimodular, we must have $\delta_i = 1$.
\\
\\
Now note that $\frac {\partial I_3} {\partial p_\theta} = 0$ and $\frac {\partial I_2} {\partial p_\theta} = -\sgn(p_\theta)$ for $g>0$ and so we may find $M_1$ because here
\begin{equation}
M_1\frac {\partial I_-} {\partial p_\theta} = \frac {\partial I_+} {\partial p_\theta}
\end{equation}
which implies that $\kappa_1 = -2$ and $\beta_1 = 0$.
\\
\\
For $g<0$ we have $\frac {\partial I_2} {\partial p_\theta} = 0$ and $\frac {\partial I_3} {\partial p_\theta} = -\sgn(p_\theta)$ and by
\begin{equation}
M_2\frac {\partial I_-} {\partial p_\theta} = \frac {\partial I_+} {\partial p_\theta}
\end{equation}
we find that $\kappa_2 = 0$ and $\beta_2 = -2$.
\\
\\
The monodromy matrix $M$ for a counter clockwise cycle around the origin $(p_\theta, g)=(0, 0)$ is given by $M = (M_2S)^{-1}(M_1S)$, which gives
\begin{equation}
M = \left( \begin{array}{ccc}
1 & 0 & 0 \\
2 & 1 & 0 \\
-2 & 0 & 1
\end{array}
\right).
\end{equation}
By a final unimodular change of basis, defined by $TMT^{-1}=N$, where an appropriate choice of $T$ is 
\begin{equation}
T = \left( \begin{array}{ccc}
1 & 0 & 0 \\
0 & -1 & -1 \\
0 & 0 & -1
\end{array}
\right),
\end{equation}
we put the monodromy matrix into normal form and have proved the following theorem:
\begin{theorem} {\em Monodromy} \label{thm:monodromy}
The obstruction to the existence of smooth global action variables for the geodesic flow on the ellipsoid with equal middle axes is monodromy.  The monodromy matrix has normal form
\begin{equation}
N = \left( \begin{array}{ccc}
1 & 0 & 0 \\
0 & 1 & 0 \\
2 & 0 & 1
\end{array}
\right). \nonumber
\end{equation}
\end{theorem}

This result is consistent with the general theory of non-degenerate singularities of integrable systems developed in \cite{matveev}, \cite{zung96a}, \cite{zung97}.  According to Nguyen Tien Zung~\cite{zung96a}, each non-degenerate singularity can topologically be presented as an almost direct product of ``basic" singularities.  In our case, this is just the direct product of the 4-dimensional focus-focus singularity with two pinched points and a non-singular system with one degree of freedom (see the last statement of theorem \ref{thm:SingFib}).  It is easily seen that the monodromy matrix in such a situation is decomposed into two blocks: $\begin{pmatrix} 1 & 2 \\ 0 & 1 \end{pmatrix}$ which correspond to the focus-focus singularity with two pinches (see \cite{matveev}, \cite{zung97}) and the trivial 1-dimensional block.  Up to a change of basis, this is exactly the matrix from theorem \ref{thm:monodromy}.

\appendix

\section{The geodesic flow on 2-ellipsoids}

Here we briefly describe the well known classical situation of
the 2-ellipsoid embedded in $\R^3$.
There are three critical periodic orbits obtained by intersecting the ellipsoid
with any coordinate plane $x_i = 0$, $i=0,1,2$. 
The middle plane $x_1 = 0$ contains 
the four umbilic points defined by $\lambda_1 = \lambda_2 = \alpha_1$ where $\lambda_i$ 
are (algebraic) elliptic coordinates on the ellipsoid. Introducing elliptic functions
$\lambda_i(\phi_i)$ gives a double covering by a torus with coordinates $\phi_i$ 
branched over the umbilic points. To analyse any motion that hits the
umbilic points these ellipsoidal covering coordinates cannot be used. 
It turns out that the only motion that ever crosses the umbilic points are the two
unstable periodic orbits in the $x_1 = 0$ plane {\em and} their separatrices.
The image of the momentum map for fixed energy is a line segment
with three corank 1 non-degenerate critical values.
The endpoints correspond to the stable orbits in the $x_0 = 0$ and $x_2 = 0$ plane.
There is another critical value in the middle of the interval corresponding to the
orbits in the plane $x_1 = 0$. The fibre of this point contains the two unstable orbits
connected by a heteroclinic separatrix. The topology is that of two circles intersecting
in two points multiplied by a circle. This can be seen by considering the Poincar\'e section
$x_0 = 0$ with $\dot x_0 = y_0 \ge 0$ on the section. The boundary of the section
$x_0 = y_0 = 0$ are itself geodesics, otherwise the flow is transverse to the section. 
The topology of the section with $\dot x_0 > 0$ is that of a finite cylinder, i.e.\ an annulus.
The restriction of the second integral $G$ to the surface of section and to constant 
energy gives a function with two minima (corresponding to the two geodesics 
in the $x_2$-plane) and two saddles (corresponding to the two geodesics 
in the $x_1$-plane). The separatrices intersect the section along two curves wrapping 
around the cylinder
once and intersecting in two points, like $p_\phi = \pm \cos \phi$.

If two semi-axes are the same we obtain a prolate or oblate ellipsoid of revolution.
For this system it is an elementary exercise to compute the non-trivial action.
For the oblate case where the longer axes are equal, $\alpha_0 < \alpha_1 = \alpha_2$ the
action is given by
\begin{equation}
I_l = \frac {1} {2\pi} \oint_{C_2} p_s ds
\end{equation}
where
\begin{equation}
p_s^2 = \left(2h - \frac {\alpha_0} {\alpha_1(\alpha_0-s^2)} J^2\right)\frac {(\alpha_0^2 + (\alpha_1-\alpha_0)s^2)} {\alpha_0(\alpha_0-s^2)}.
\end{equation}
The essential integral is a function of the ratio $\rho = \alpha_0/\alpha_1$ and the scaled angular 
momentum $\hat j = j / \sqrt{2h \alpha_1}$ only. 
Expressing $I_l$ in Legendre normal form gives
\begin{equation}
\frac 14 \frac{2\pi I_l}{\sqrt{2h\alpha_1}} = 
U {\cal E}(k) - \frac{\rho \hat j}{U} \Pi ( \beta^2, k),
\quad 
U^2 = 1 - \hat j^2( 1 -\rho )\, , 
\end{equation}
where ${\cal E}$ and $\Pi$ are Legendre's complete elliptic integrals of the second and third kind respectively and the modulus and parameter are
\begin{equation}
k^2 = 1 - \rho  U^{-2}, 
\qquad 
\beta^2 = {k^2}/ {1 - \rho}\, .
\end{equation}
For the oblate case $\rho > 1$ and thus $k^2 < 0$. Upon replacing $U$ by $\imag U$ 
the above formula for the action also holds in this case.

\begin{figure}
\centerline{\includegraphics[width=12cm]{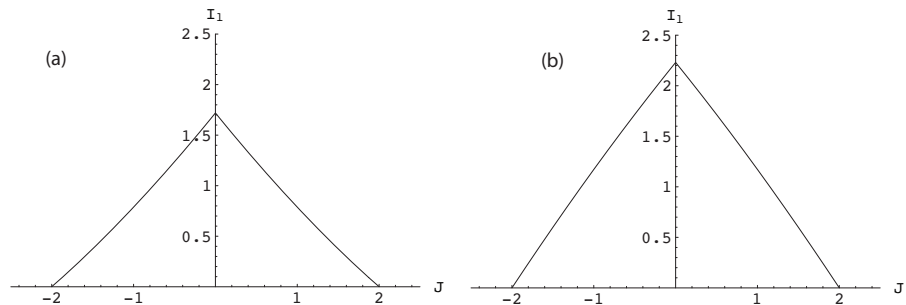}}
\caption{Actions for ellipsoids of revolution: (i) $\alpha_0<\alpha_1=\alpha_2$, (ii) $\alpha_1=\alpha_2<\alpha_3$, where $\alpha_0=1$, $\alpha_1=2$, $\alpha_3=4$, $h=1$.}
\label{fig:fig2dacts}
\end{figure}

\bibliographystyle{plain}
\bibliography{PaperBibfile}

\end{document}